\documentclass[%
 reprint,
superscriptaddress,
 nofootinbib,
 amsmath,amssymb,
 aps,
pra,
]{revtex4-1}

\usepackage[utf8]{inputenc}
\usepackage{graphicx}
\usepackage{epstopdf}
\usepackage{braket}
\DeclareGraphicsExtensions{.eps,.pdf}

\usepackage{dcolumn}
\usepackage{bm}
\usepackage{siunitx}

\DeclareSIUnit\gauss{G}
\usepackage{mathrsfs,amsmath}
\usepackage{siunitx}
\usepackage{pgfplots}
\usepackage{pgfplotstable,filecontents}
\usepackage{hyperref}
\hypersetup{
  colorlinks=true,
  linkcolor=blue!50!red,
  urlcolor=green!50!black,
	citecolor=blue!50!blue
}

\pgfplotsset{compat=1.9}

\begin{document}

\preprint{APS/123-QED}

\title{Coherent conversion between microwave and optical photons - an overview of physical implementations}

\author{Nicholas J.\ Lambert}
\affiliation{Department of Physics, University of Otago, Dunedin, New Zealand}
\affiliation{The Dodd-Walls Centre for Photonic and Quantum Technologies, New Zealand}
\author{Alfredo Rueda}
\affiliation{The Dodd-Walls Centre for Photonic and Quantum Technologies, Dunedin, New Zealand}
\affiliation{Institute of Science and Technology Austria, am Campus 1, 3600 Klosterneuburg, Austria}%
\author{Florian Sedlmeir}
\affiliation{Max Planck Institute for the Science of Light, Staudtstr. 2, 90158 Erlangen, Germany }%
\affiliation{Institute for Optics, Information and Photonics, University Erlangen-Nuernberg, Staudtstr. 7/B2, 91058 Erlangen, Germany}
\author{Harald G.\ L.\ Schwefel}
\email{Harald.Schwefel@otago.ac.nz}
\affiliation{Department of Physics, University of Otago, Dunedin, New Zealand}
\affiliation{The Dodd-Walls Centre for Photonic and Quantum Technologies, New Zealand}
\date{\today}

\begin{abstract}
Quantum information technology based on solid state qubits has created much interest in converting quantum states from the microwave to the optical domain. Optical photons, unlike microwave photons, can be transmitted by fiber, making them suitable for long distance quantum communication. Moreover, the optical domain offers access to a large set of very well developed quantum optical tools, such as highly efficient single-photon detectors and long-lived quantum memories. For a high fidelity microwave to optical transducer, efficient conversion at single photon level and low added noise is needed.  Currently, the most promising approaches to build such systems are based on second order nonlinear phenomena such as optomechanical and electro-optic interactions. Alternative approaches, although not yet as efficient, include magneto-optical coupling and schemes based on isolated quantum systems like atoms, ions or quantum dots. In this Progress Report, we provide the necessary theoretical foundations for the most important microwave-to-optical conversion experiments, describe their implementations and discuss current limitations and future prospects.
\end{abstract}

\pacs{Valid PACS appear here}

\maketitle
\tableofcontents

\section{Introduction}

The building block for quantum information technologies is the qubit - a two level system the quantum state of which can be prepared, manipulated and measured~\cite{DiVincenzo2000}. Many promising solid state implementations of qubits have been demonstrated, including superconducting qubits of diverse flavours~\cite{Devoret1169,Wendin_2017}, spin qubits~\cite{PhysRevA.57.120,Vandersypen2017}, and charge qubits in various material systems~\cite{PhysRevLett.91.226804,Pashkin2009}. Typical characteristic energy scales for these systems correspond to radiation frequencies of order $\SI{10}{\giga\hertz}$, allowing them to be readily manipulated using commercial microwave technology.

Proposals for novel quantum information processing techniques often rely on a quantum network~\cite{Kimble2008,PhysRevLett.78.3221,Sasaki_2017,Komar2014,Kurpiers2018}, linking together multiple qubits or groups of qubits to enable quantum-secure communication~\cite{PhysRevA.68.042331, Grosshans2003}, novel metrology techniques~\cite{Kok2004,Escher2011}, or distributed quantum computing~\cite{Beals2013}. However, microwave frequency photons are difficult to transmit over long distances -typical attenuation in low-loss microwave cables at \SI{10}{\giga\hertz} is more than \SI{1}{\decibel\per\metre}, which compares very poorly with optical fibres with losses below \SI{0.2}{\decibel\per\kilo\metre} at telecom wavelengths ($\lambda\approx\SI{1550}{\nano\metre}$, $f\approx\SI{193}{\tera\hertz}$). The advantages of transmitting quantum information over fibers is immediately apparent.

Furthermore, the thermal occupancy of optical frequency channels is close to zero at room temperature, in contrast to microwave modes; a mode with a frequency of \SI{10}{\giga\hertz}  must be cooled below \SI{100}{\milli\kelvin} to reduce the average photon occupancy below 1\%. The availability of telecom band single photon detectors, quantum memories~\cite{jevonmemory,lvovsky_optical_2009} and other technologies common in quantum optics experiments~\cite{bagci_optical_2014} also suggests the need for the development of techniques for the bi-directional transfer of quantum information between microwave and optical photons.

Such a transducer must have a high fidelity, or quantum capacity. Although error correcting quantum algorithms exist~\cite{doi:10.1098/rsta.2011.0494,Campbell2017}, they typically require an error rate less than 1\%~\cite{PhysRevA.80.052312} with the precise figure depending on both the scheme to be implemented and the nature of the errors. A transducer must also have a high quantum efficiency - close to one output photon must be produced for every input photon. Quantum capacity is finite only if the conversion efficiency is greater than 50\%~\cite{PhysRevLett.98.130501}, although indirect schemes involving heralded entanglement of photons may avoid this limit~\cite{arXiv:1901.08228}.

Qubit implementations relying on microwave excitations are typically operated in dilution fridges with base temperatures of around \SI{10}{\milli\kelvin}. The transducer must therefore operate in a cryogenic environment. This also prevents inadvertent up-conversion of thermal microwave photons, but places stringent restrictions on the power dissipation in the device; the cooling power of a dilution fridge at \SI{100}{\milli\kelvin} is typically only a few hundred microwatts. 

Finally, quantum systems rapidly lose information to their environment due to decoherence. The device must therefore have enough bandwidth for sufficient information to be transmitted before it is lost - the best decoherence times for superconducting qubits approach \SI{0.1}{\milli\second}~\cite{PhysRevB.86.100506}, corresponding to a bandwidth of \SI{10}{\kilo\hertz}.

These requirements combine to make the task a challenging one. Efficient frequency mixing cannot occur unless a significant non-linearity is introduced. This can come from the susceptibility of a transparent material such as lithium niobate (LiNbO$_3$), leading to an electro-optic non-linearity. Alternatively, more extreme non-linearities are found near the resonances of three (or more) level systems, such as rare earth ions in crystals, or rubidium vapors. The non-linearity can also emerge due to indirect coupling mediated by another mode, such as mechanical vibrations or magnetostatic modes.

The effect of non-linearities can be increased by placing the material in resonant cavities, where they experience both an enhanced photon interaction time, and a modified density of optical states. Because of the importance of cavities to experimental implementations of microwave-to-optical transducers, we start this Progress Report by giving an overview of the physics of cavity modes. We then detail current experimental approaches, before summarizing progress to date and outlining possible future directions.

{
\setlength{\tabcolsep}{1em}

\begin{table*}
\caption{Symbols and their meanings.\\}
\begin{tabular}{ c c c }
\hline
Symbol&Meaning&Units\\
\hline
\hline
$\omega$ & 				 Angular frequency &																   \si{\per\second} \\
$\omega_n$ &	     Angular frequency of optical mode $n$ &                       \si{\per\second} \\
$\Omega$ &         Angular frequency of microwave signal &               \si{\per\second} \\
$\kappa'_n$ &     Dissipative loss rate of cavity mode  intensity (for mode $n$) &                 \si{\per\second} \\
$\kappa_{n,(j)}$ &     		External loss rate of cavity mode intensity at $j$th port (for mode $n$) &                 \si{\per\second} \\
$\kappa_{\textrm{e},n}$ &     		Total external loss rate of cavity mode intensity (for mode $n$) &                 \si{\per\second} \\
$\kappa_n$ &     		Total loss rate of cavity mode intensity (for mode $n$)  &                 \si{\per\second} \\
$\gamma'_n$ &     Dissipative loss rate of cavity mode field (for mode $n$) &                 \si{\per\second} \\
$\gamma_{n,(j)}$ &   External loss rate of cavity mode field at $j$th port (for mode $n$) &    \si{\per\second} \\
$\gamma_{\textrm{e},n}$ &   Total external loss rates of cavity mode field (for mode $n$) &    \si{\per\second} \\
$\Gamma_n$ &         Total loss rate of cavity mode (for mode $n$) &		                   \si{\per\second} \\
$Q$ &              Quality factor of cavity mode $n$, $\frac{\omega_n}{2 \Gamma_n}$ &    1 \\
$g$ &						   Coupling rate &                                       \si{\per\second} \\
$\rho(\omega)$ &	 Local density of optical states &										 \si{\joule\per\cubic\meter} \\
$\hat{a}^\dagger$, $\hat{a}$ & Creation and annihilation operators for field $a$ &   1 \\
$\chi^{(n)}$ &        $n$-th order electric susceptibility &                1 \\
$\eta$ &           Quantum efficiency	&																	 1 \\
$T_i$ &           Internal temperature of cavity	&											\si{\kelvin} \\
$T_j$ &           Temperature at $j$th port of cavity	&									\si{\kelvin} \\

\hline
\hline
\end{tabular}
\end{table*}
}

\section{Cavities}

\subsection{Cavity properties}

\label{cavityproperties}

Electromagnetic cavities\footnote{The terms \textit{resonator} and \textit{cavity} are sometimes used interchangeably in the literature. Here, for brevity, we use the term cavity to describe all structures supporting an electromagnetic mode, including metallic cavities, transmission line resonators, Fabry-P\'erot type optical cavities formed from parallel mirrors, photonic crystal cavities, and dielectric resonators.} support long-lived localized electromagnetic modes, characterized by a resonant frequency $f_0$. The mode is subject to loss, which might be due to radiative losses, absorption by scattering centers, ohmic losses, dielectric losses, etc; these are termed dissipative losses, and we denote the dissipative intensity loss rate due to all these effects by $\kappa '$. (Alternatively, the field loss rate $\gamma '=\kappa '/2$ can be used.) Typically the mode is probed via its coupling to travelling waves that are propagating either in free space, or in waveguides such as a coax cable or optical fibre, through one or more ports with coupling rates $\kappa_j$. The mode can be excited via these ports, but also loses energy through them. The total loss rate is therefore the sum of dissipative losses, and external losses through ports, $\kappa = \kappa'+\sum_j\kappa_j$. This gives the linewidth of the mode. Narrow linewidths correspond to long photon confinement times $\tau_p = 1/\kappa$, allowing longer interaction lengths for weak non-linear effects to become significant. The quality factor of a mode of angular frequency $\omega_0$ is defined as $Q = \omega_0\tau_p$, with a higher quality factor ($Q$) corresponding to a longer lifetime for photons in the cavity at $\omega_0$, but also a slower response of the system to a change in stimulus and hence a smaller bandwidth.

\begin{figure}
\includegraphics{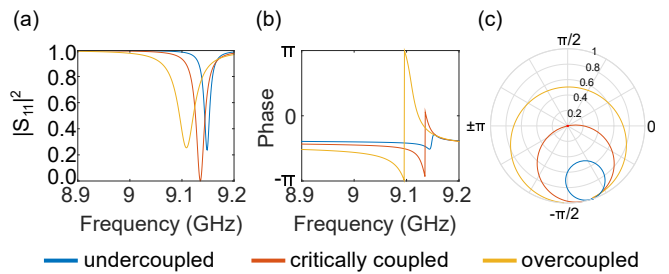}
\caption{Simulated $S_{11}$ for an under-, over- and critically coupled 3D cavity. (a) Amplitude of reflected wave. At the resonant frequency $\Omega$, the reflected power is finite for over- and under-coupling, but goes to zero for critical coupling. The shift of $\Omega$ to lower frequencies due to increasing coupling is also apparent. (b) Phase of reflected power. The initial phase in these simulation is arbitrary, and the jump in the phase is due to the discontinuity at $\phi=\pm\pi$. (c) The same data plotted parametrically on polar axes. Distinguishing over-coupled (circle encloses the origin), critically coupled (circle goes through the origin) and under-coupled (circle does not enclose the origin) measurements is now straightforward.}\label{fig:over_under}
\end{figure}

The cavity mode can be characterized by exciting the mode via a port, and measuring the reflected power. The ratio of reflected to incident power is termed $\textrm{S}_{11}$. (Modes can also be probed in transmission by measuring the power emitted from a second port, but this is undesirable because the absence of a baseline makes analysis of resonator loss impossible.) The interplay between the coupling rate at the measured port $\kappa_1$ and the other loss rates $\kappa'+\sum_{j\geq2}\kappa_j$ allows three different regimes to be defined; under-coupled ($\kappa_1 < \kappa'+\sum_{j\geq2}\kappa_j$), over-coupled ($\kappa_1 > \kappa'+\sum_{j\geq2}\kappa_j$), and critically coupled ($\kappa_1 = \kappa'+\sum_{j\geq2}\kappa_j$). They can be distinguished by examining both real and imaginary parts (or, equivalently, amplitude and phase) of the reflected signal as a function of frequency. Typical data for a 3D microwave cavity are shown in Fig \ref{fig:over_under}, and the analysis can be made robust against noise and reflections by including terms describing the environment to which it is coupled~\cite{doi:10.1063/1.4907935}.

Besides coherent excitation at the input port, at finite temperatures a cavity mode is thermally occupied. For a mode of temperature $T$ and angular frequency $\omega$, the mean thermal occupancy is $n_\textrm{th}=k_\textrm{B}T/\hbar\omega$. The temperature of the mode can be reduced by reducing the temperature of the internal cavity environment, $T_i$, but is also dependent on the external temperatures, $T_j$, to which the mode is coupled via the $j$th port. The mode temperature is given by the weighted average of all coupled temperatures,
\begin{equation}
T=\frac{\kappa'}{\kappa}T_i + \sum_j\frac{\kappa_{(j)}}{\kappa}T_{(j)}.
\end{equation}
Thermal occupancy is negligible for optical frequency modes even at room temperature ($n_\textrm{th} < 10^{-13}$), but is significant for microwave modes ($\Omega \sim \SI{10}{\giga\hertz}$ unless cooled to cryogenic temperatures ($n_\textrm{th} = 10^{-2}$ at \SI{104}{\milli\kelvin}. These photons can be a significant source of noise for microwave up-conversion, particularly at the single photon limit, where incidental up-conversion of thermal photons can dominate if they are not suppressed.

It is also useful to define the volume of the cavity mode, but this can be difficult for cavities with finite coupling to the environment, with the situation being particularly troublesome for open cavities. In early literature~\cite{Purcell1946}, the physical volume of the cavity, $V$ was used for the mode volume. This is a somewhat crude estimation and
\begin{equation}
V=\frac{\int_V \epsilon(r) | \mathbf{E(r)}|^2}{(\epsilon(r) | \mathbf{E(r)}|^2)_\textrm{max}}
\end{equation}
for electric field $\mathbf{E(r)}$ is generally found to give a better value for the volume of the electric field of the mode~\cite{LesHouches1990}, particularly when the spatial boundaries of the mode are clearly defined.  But the integral diverges if allowed to run over all space~\cite{PhysRevLett.110.237401}, and for very leaky cavities it is not always obvious what the best renormalization approach is~\cite{Kristensen:12, PhysRevB.94.235438, PhysRevA.92.053810, PhysRevA.96.017801, PhysRevA.96.017802, PhysRevA.49.3057}. (If we care about the magnetic component of the cavity field rather than the electric component, $\mathbf{E(r)}$ is replaced by $\mathbf{B(r)}$ and $\epsilon(r)$ is replaced by $1/\mu(r)$.)

\subsection{Coupled systems}

The coherent interaction between photons and other systems is at the heart of many technologies, but for photons in travelling waves it is frequently too weak to be useful. To enhance the interaction time and strength, the photons are confined to a cavity. Using this approach, the interaction can be increased such that the states of two systems are hybridized, and they cannot be described separately. Instead, they are described by cavity quantum electrodynamics, and applications of this theory have shown their usefulness for the control of quantum systems such as atoms, qubits and spin ensembles~\cite{Mabuchi1372,Walther2006}.

The coupling strength associated with the exchange of an energy quantum between the two interacting systems is dependent on the overlap between the final state and the single photon Hamiltonian acting on the initial state. It is given by the transition matrix element
\begin{equation}
g=\frac{1}{\hbar}\bra{f}\hat{H}_\textrm{int}\ket{i}, 
\end{equation}
where $i$ ($f$) is the state of the system before (after) the exchange. Here, $g$ is the single photon coupling strength, and it is very often desirable to make this as large as possible. This can be done by increasing the magnitude of the dipole transition linking the two states, co-aligning the dipole and the field, or increasing the strength of the single-photon electric or magnetic field in the cavity at the location of the dipole. The strength of the field at the dipole can be changed by varying the position of the dipole in the cavity; positioning the resonator at a cavity field antinode maximizes the coupling. The strength of the field can also be increased by reducing the electric or magnetic volume of the cavity, and therefore increasing the confinement of the photon~\cite{PhysRevA.92.053810}. The relative volumes of the magnetic and electric parts of the mode's electromagnetic field are characterized by its impedance. By tailoring the mode form, it can be chosen to be either high impedance, to maximize the coupling to electric dipole moments~\cite{PhysRevB.97.235409,PhysRevB.95.224515,PhysRevX.7.011030,doi:10.1002/andp.200710261}, or low impedance, maximizing coupling to magnetic dipole moments~\cite{PhysRevA.95.022306,PhysRevLett.118.037701,PhysRevB.99.140414}. 

\subsection{Weak and strong coupling}

Two coupling regimes can be identified; in the weak coupling regime, the coupling strength is less than the linewidths of the two resonances. The interaction can then be treated as a second order perturbation on each system due to the other, leading to a change in $\omega_1$ and the spontaneous decay rate. The latter phenomenon is termed the Purcell effect~\cite{Purcell1946}, and is due to the cavity introducing a frequency dependency to the local density of optical states (LDOS) $\rho(\omega)$. For a cavity supporting several modes labelled with index $n$
\begin{equation}
\rho(\omega)=\sum_n\frac{1}{\pi}\frac{\Gamma_{n}/2}{(\omega-\omega_{n})^2+\Gamma_{n}^2/4},
\end{equation}
which is the product of the LDOS of free space with a sum of Lorentzian lineshapes of widths $\Gamma_n$ and frequencies $\omega_n$. The transition rate, $\Gamma_{i \rightarrow f}$, of a two level system coupled to the cavity is now given by Fermi's Golden rule,
\begin{equation}
\Gamma_{i \rightarrow f}= \frac{2 \pi} {\hbar}  \left | \bra{f}\hat{H}_\text{int}\ket{i}\right |^{2} \rho(\omega).
\end{equation}

The effect of the cavity is therefore to enhance the spontaneous emission rate at frequencies close to resonance, and suppress it away from resonance. The enhancement on resonance over the emission rate in free space, $\Gamma_\mathrm{free}$,  is termed the Purcell factor, and is given by
\begin{equation}
P=\frac{\Gamma_{i \rightarrow f}}{\Gamma_\mathrm{free}}=\frac{3}{4\pi^2}\left(\frac{\lambda_\mathrm{free}}{n}\right)^3\left(\frac{Q}{V}\right).
\end{equation}
The requirements for a large Purcell factor follow those for enhanced coupling in so far as a small cavity volume enhances the relaxation rate, but the LDOS is also enhanced by increasing the mode lifetime. Control of the LDOS by cavities protects qubits from spontaneous emission~\cite{PhysRevApplied.7.054020,PhysRevLett.112.190504,PhysRevLett.106.030502,Bronn2015}, controls relaxation in spin ensembles~\cite{Bienfait2016} and tunes the emission properties of single photon sources~\cite{PhysRevLett.95.013904, PhysRevLett.96.117401, doi:10.1063/1.2189747, Liu2018, Gallego:18}.

Alternatively, if $g$ is larger than the linewidths, the system is in the strong coupling regime, in which the coupling rate is larger than the loss rates, and a perturbative analysis does not apply when the two are close to resonance. In particular, if $\omega_{a} = \omega_{b}$, the degeneracy is lifted by the interaction, and a doublet of new eigenstates of eigenfrequencies $\omega_{a}\pm \frac{1}{2}g$ are formed. This effect is termed Rabi splitting, and the system can be described in terms of light-dressed atom states. The Hamiltonian of the system is given by the Rabi equation,
\begin{align}
\hat{H} &= \hat{H}_a +\hat{H}_b + \hat{H}_\textrm{int}\\
&=\hbar \omega_{a} (\hat{a}^\dagger \hat{a})+\hbar \omega_{b}(\hat{b}^\dagger \hat{b}) + \frac{\hbar g}{2}(\hat{a}^\dagger + \hat{a})(\hat{b}^\dagger + \hat{b}).
\end{align}
Here, $\hat{a}^\dagger$ ($\hat{a}$) is the raising (lowering) operator for the first cavity, and $\hat{b}^\dagger$ ($\hat{b}$) is the raising (lowering) operator for the second cavity or two level system. By expanding the interaction term and ignoring the high frequency components (the \textit{rotating wave approximation}) we arrive at the beam splitter Hamiltonian.
\begin{equation}\label{beamsplitter}
\hat{H} = \hbar \omega_{a} (\hat{a}^\dagger \hat{a})+\hbar \omega_{b}(\hat{b}^\dagger \hat{b}) + \frac{\hbar g}{2}(\hat{a} \hat{b}^\dagger + \hat{a}^\dagger \hat{b}).
\end{equation}
The interaction term now describes the coherent transfer of energy between the two systems. The rotating wave approximation is valid only when $g\ll\omega_1,\omega_2$. If this is not the case, then the system is in the ultrastrong coupling or deep strong coupling regimes~\cite{arXiv:1804.09275, Kockum2019}.

Whilst the above discussion is framed in terms of photons in electromagnetic cavities, any other oscillating system that admits second quantization can be treated in the same way. In particular, mechanical vibrations in high $Q$ membranes and beams can be described in terms of phonons, and excitations of magnetostatic modes as magnons. Equation (\ref{beamsplitter}) describes coherent conversion between bosonic modes; if $\hat{b}^\dagger$($\hat{b}$) is replaced with the atomic raising (lowering) operator $\hat{\sigma}_+$($\hat{\sigma}_-$), the equation describes coherent conversion between boson modes and two level systems, and is known as the Jaynes-Cummings Hamiltonian.

The objective of up-conversion is to exploit these Hamiltonians to achieve coherent conversion between a microwave frequency mode and an optical mode, perhaps via other modes. In the following sections, we examine specific systems in which this can be implemented.

\section{Experimental approaches}

We now describe the state of the experimental art of microwave upconversion. We divide the approaches into those that rely on a non-linear electro-optic coupling, those requiring a non-linear magneto-optic coupling, those that are best described as multi-level systems, and those where interactions between photons is mediated by a mechanical element.

\subsection{Electro-optic coupling}

\begin{figure}
\includegraphics{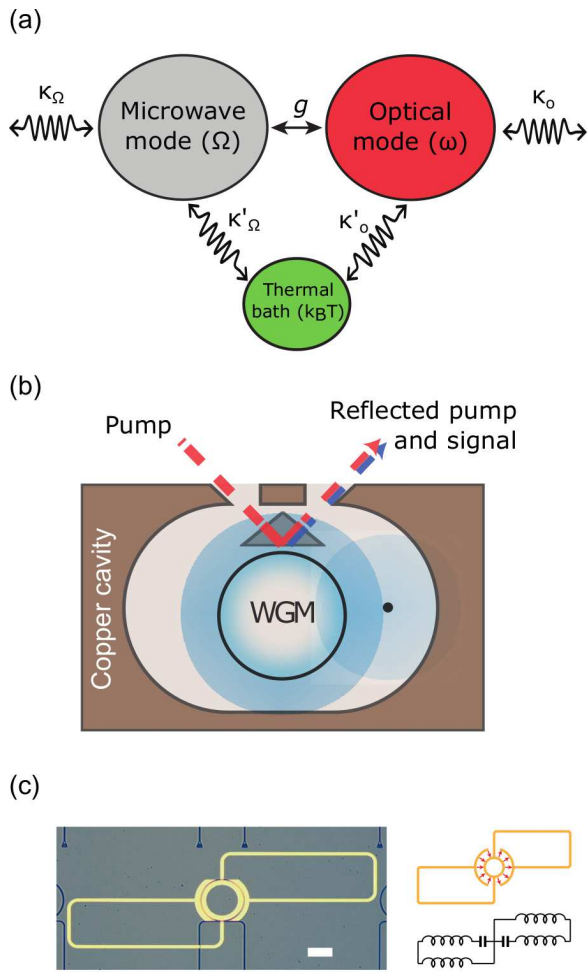}
\caption{Electro-optic upconversion. a) Cartoon of process, showing coupling $g$ between optical and microwave modes due to electro-optic non-linearity, and noise introduced by a thermal bath. (b) Realization using a 3D copper cavity and a lithium niobate WGM resonator, from Ref.~\protect{\cite{Rueda:16}}. Light is prism-coupled in and out of the WGMs. (c) Realization using a coplanar superconducting cavity and an AlN ring resonator, from Ref.~\protect{\cite{Fan:18}}.}\label{fig:electrooptic}
\end{figure} 

Alternating electric fields within a crystal displace charges from their zero-field equilibrium sites, leading to bond rotation and bond stretch, and oscillations with harmonic and anharmonic components~\cite{0022-3719-15-4-027,PhysRevLett.44.281}. The effect is described by the material's susceptibility $\chi(\omega,\mathbf{E})$, which characterizes its polarization in response to an external electric field $\mathbf{E}$. Very often, a linear approximation can be made, and this becomes simply $\chi(\omega)$, embodied by the refractive index. But in media with significant anharmonic components, this is not the case~\cite{boyd2003nonlinear}. The tensor electric susceptibility can be written as 
\begin{equation}
\mathbf{P}(t) = \varepsilon_0( \chi^{(1)} \mathbf{E}(t) + \chi^{(2)} \mathbf{E}^2(t) + \chi^{(3)} \mathbf{E}^3(t) + \ldots),
\end{equation}
where $\chi^{(n)}$ is the $n$-th order electric susceptibility, and is a tensor of rank $(n+1)$. The presence of a non-zero $\chi^{(2)}$  is only possible in crystals lacking inversion symmetry, while the $\chi^{(3)}$ term does not require a special symmetry and is present in amorphous media such as liquids and glasses.

A large $\chi^{(2)}$, as well as leading to the electro-optic Pockels effect, allows different frequency fields to interact. Three wave interactions, such as parametric down conversion, sum frequency generation (SFG) and difference frequency generation (DFG) (also termed anti-Stokes and Stokes processes respectively) are made possible. SFG, is the combination of waves at frequencies $\omega_1$ and $\omega_2$ to create a new wave at $\omega=\omega_1+\omega_2$. DFG, on the other hand, is the generation of a wave at $\omega=\omega_1-\omega_2$ from waves at $\omega_1$ and $\omega_2$. This process must also result in additional power at $\omega_2$ due to conservation of energy.

DFG can occur spontaneously, creating an incoherent microwave population and therefore adding noise. This process can be useful; use of it has been proposed to generate entangled pairs of microwave and optical photons~\cite{tsang_cavity_2011}. Above threshold, the spontaneous process can stimulate parametric oscillations generating coherent microwave radiation~\cite{Savchenkov:07}. But for coherent up-conversion, DFG is undesirable.

In general, the non-linearities in transparent materials far from resonance are small, and intense fields, such as those produced by lasers, are required to observe significant effects. Electromagnetic cavities are a natural choice to enhance the electric and magnetic fields in the medium, as well as providing ways to manipulate energy levels in the device with the toolkit of quantum electrodynamics.

Non-linear electro-optic materials with a significant second order non-linear polarizability $\chi^{(2)}$ allow microwave frequencies to be used to modulate the phase and intensity of incident light through the Pockels effect. This effect is used in commercial electro-optic modulators, and can be used to generate SFG and DFG sidebands. Indeed, a commercial electro-optic modulator~\cite{ThorLabs} can be used for microwave upconversion with an efficiency of $\eta \approx 3\times 10^{-7}$. In Ref.~\cite{Khan:07} a GaAs crystal was used to observe the up-conversion of radiation from \SI{700}{\giga\hertz} to telecom wavelengths. The efficiency was measured to be $\eta=10^{-5}$ with no resonant enhancement for either the optical or microwave fields.

\begin{figure}
\includegraphics{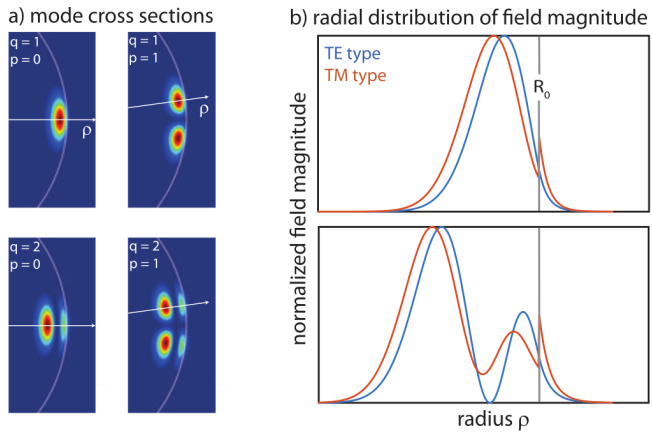}
\caption{Spatial form of WGMs. (a) Scalar field distribution for TE type modes. The $q = 1$, $p = 0$ is the fundamental mode, while $p > 0$ and $q > 1$ are of higher polar and radial order respectively. (b) shows the scalar field distribution as function of the radius for TE and TM modes along the white arrows in (a). Here, the different boundary conditions for TE and TM polarization become apparent: TE is continuous and TM has a discontinuity at the boundary ($\rho = R_0$). Figure from Ref.~\protect{\cite{Sedlmeir2016}}.}\label{fig:wgms}
\end{figure}

A large $\chi^{(2)}$ is also found in LiNbO$_{3}$, and this was first used for up-conversion of lower energy microwaves with a frequency of \SI{100}{\giga\hertz} by Strekalov et al.~\cite{ilchenko_whispering-gallery-mode_2003,strekalov_efficient_2009}. In order to increase the efficiency, the optical pump field was confined in high-quality ($Q \sim 10^6$) whispering gallery modes (WGMs)~\cite{strekalov_microwave_2009} supported in a LiNbO$_{3}$ disc. WGMs are formed when light propagates along a closed loop formed by a step change in refractive index. Typically this is either a circular disc or microdisc, or a spheroid of a dielectric material~\cite{reviewdmitry} in air or vacuum. Modes are denoted by angular ($m$), radial ($q$) and polar ($p$) indices, and both transverse--electric (TE) and transverse--magnetic (TM) mode families are supported. Examples of spatial mode forms are shown in Fig.~\ref{fig:wgms}.

For modes with fixed radial and polar numbers in a circular disk of radius $r$, the frequency separation between successive angular modes (the free spectral range, FSR) is given by $c/2 \pi r n_{\textit{eff,mqp}}$. Here, $n_{\textit{eff,mqp}}$ is an effective refractive index lower than that of the dielectric, due to the mode field lying partly outside the disk, and is dependent on the polarization of the mode as well as the angular, radial and polar indices. For dispersive media, it is also frequency dependent. For a FSR of \SI{9}{\giga\hertz}, a LiNbO$_3$ disc must have a radius of approximately \SI{2.5}{\milli\metre}. Coupling light from free space to the mode is achieved through the evanescent field, but this cannot be achieved directly due to the index mismatch between the two media leading to a phase mismatch between the incoming field and the confined mode. Either waveguide or prism coupling is used; in~\cite{strekalov_microwave_2009} a diamond prism was used, allowing close-to-critical coupling to be achieved. By using a birefringent coupling prism (for example rutile or LiNbO$_3$), TE and TM modes can be out-coupled along spatially separated paths. Alternatively, the beams can be separated with the use of a polarizing beam splitter~\cite{PhysRevApplied.9.024007}. 

\begin{figure}
\includegraphics{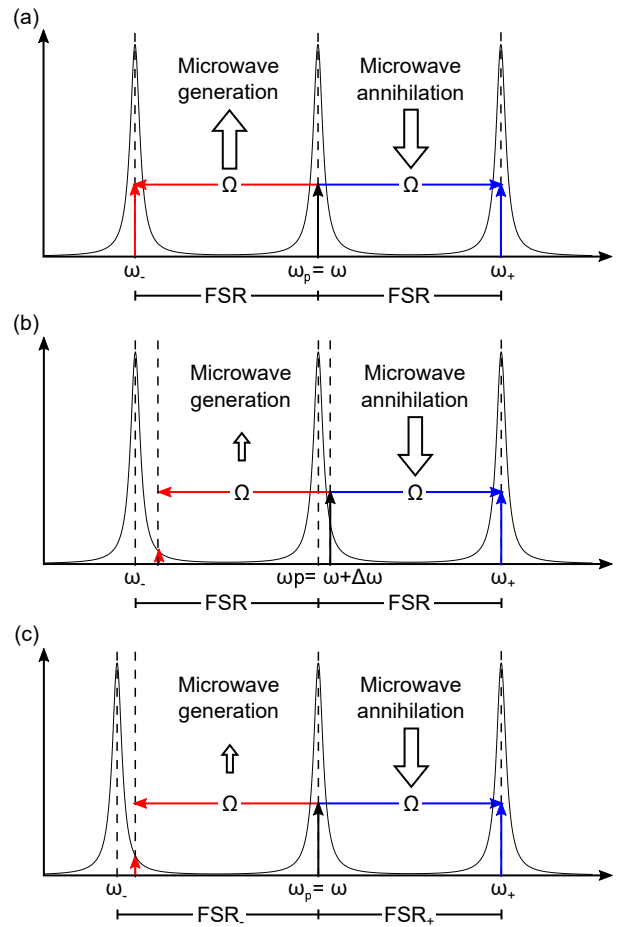}
\caption{The triple resonance condition. a) The frequency of the microwave mode must match the spacing between optical modes for efficient frequency conversion. However, a uniform FSR also leads to efficient \textit{down}-conversion. The down-conversion can be supressed by either (b) detuning the optical pump or (c) exploiting a non-uniform FSR.}\label{fig:tripleres}
\end{figure}

For efficient microwave up-conversion the triple-resonance condition (Fig.~\ref{fig:tripleres}), in which the pump, microwave signal, and optical signal are all resonant with electromagnetic modes, must be met, and so both the output frequency and the input frequency must coincide with that of an optical mode. If this is not the case, the rate of the up-conversion process will be heavily suppressed due to the decreased LDOS. This places a constraint on the allowed microwave frequencies; they must be an integer multiple of the FSR.

In Ref.~\cite{strekalov_efficient_2009} both Stokes and anti-Stokes processes were observed, but a small detuning of the microwave frequency resulted in the anti-Stokes process being preferential. This was due to the presence of dispersion in the LiNbO$_3$ disk, leading to a non-uniform FSR. Tuning of the frequency separation of the WGMs, and thus selection of SFG over DFG, is of critical importance for efficient and noise-free up-conversion. FSR tuning has been achieved in the context of resonant frequency combs by engineering the dispersion of the WGMs by shaping the edge of the disk~\cite{Ilchenko:03,Yang:16,Zhou:17}. In that case, however, the aim was rather to minimize dispersion and thus make the FSR independent of frequency.

A further increase in efficiency can be obtained by resonant enhancement of the microwave field (Fig.~\ref{fig:electrooptic}a). In Ref.~\cite{Rueda:16}, a LiNbO$_3$ disk was embedded in a solid copper cavity (Fig.~\ref{fig:electrooptic}b), tuned to match the FSR = \SI{8.9}{\giga\hertz} of the WGMs. The loaded microwave cavity had a quality of $Q\approx200$ at room temperature, limited by ohmic losses; the carefully polished LiNbO$_3$, however, showed a particularly high optical $Q$ of $\sim 10^8$. As well as up-conversion, the system was also used to demonstrate efficient optical comb generation~\cite{Rueda19}.

In a double resonant system, the key parameter controlling conversion efficiency is the multiphoton cooperativity~\cite{tsang_cavity_2010,tsang_cavity_2011}, $G_0 = n_pg^2/\Gamma_{(o)} \Gamma_{(\Omega)}$. Here, $n_p$ is the total photon number of the optical pump, and $g$ is the non-linear coupling between the two modes, given by
\begin{equation}
g=2 \epsilon_0 \chi^{(2)}\sqrt\frac{\hbar \omega_{1} \omega_{2} \omega_{3}}{8\epsilon_1\epsilon_2\epsilon_3 V_1 V_2 V_3}\int{dV \psi_3^\dagger \psi_2^{\phantom{\dagger}} \psi_1^{\phantom{\dagger}}}.
\label{eog}
\end{equation}
$\chi^{(2)}$ (the electrooptic nonlinear coefficient) and $\epsilon_{1,2,3}$ (the permittivities at microwave, pump and signal mode frequencies $\omega_{1,2,3}$) are material dependent parameters. The integral gives the spatial overlap between modes 1, 2 and 3 with field distributions $\psi_{1,2,3}$. $V_{1,2,3}$ are the mode volumes, defined by $V_i=\int dV \psi_i^*\psi_i^{\phantom{\dagger}}$. The overlap should be maximized to increase $G_0$, while the mode volumes, on the other hand, should be minimized.

The conversion efficiency $\eta={N^{\text{output}}_b}/{N^{\text{input}}_\Omega}$ is given by~\cite{Fan:18, RuedaThesis}
\begin{equation}
\eta=\frac{4\gamma_{\textrm{e},o}\gamma_{\textrm{e},\Omega}}{\Gamma_{o}\Gamma_{\Omega}}\frac{G_0}{(1+G_0-\frac{\Delta\Omega^2}{\Gamma_o\Gamma_\Omega})^2+\frac{\Delta\Omega^2}{\Gamma^2_o\Gamma^2_\Omega}(\Gamma_o+\Gamma_\Omega)^2}, \label{conversiontraveling}
\end{equation}
with $\Gamma_{o(\Omega)}$ and $\gamma_{\textrm{e},o(\Omega)}$ being the total and external field loss rates of the output (microwave) modes, and $\Delta\Omega$ the detuning of the microwave signal from the microwave mode. For unity conversion efficiency to be possible, $G_0\ge 1$, with further increases broadening the achievable bandwidth~\cite{tsang_cavity_2011} but requiring higher optical pump power.

To maximize the overlap between WGMs and microwave modes, the microwave field must be focused upon the edge of the disk, or the equator of the sphere, where the optical WGMs lie. For rotationally symmetric modes the integral in Eqn (\ref{eog}) leads to Clebsch-Gordan like selection rules that directly lead to a form of angular momentum conservation. In nonlinear optics this is also known as phase matching, which intuitively requires the phase velocity of the interacting fields to coincide. In Ref.~\cite{Rueda:16}, the microwave mode was confined by a pair of toroidal pillars with the same radius as the disk, but because a standing wave was excited in the cavity, half of the microwave photons were lost to the counter-propagating mode.

In order to reduce Stokes processes, the system in Ref.~\cite{Rueda:16} was thermally tuned to an anticrossing between the bright TE modes and the dark TM modes. This resulted in sufficient asymmetry in the spacing of the adjacent WGMs that emission into the lower sideband was completely suppressed (Fig.~\ref{fig:tripleres} (c)). A cooperativity of $G_0 \approx 4 \times10^{-3}$ was demonstrated, leading to an efficiency of $\eta = (1.09\pm0.02) \times 10^{-3}$.

Up-conversion in an aluminum nitride micro-ring cavity of diameter \SI{200}{\micro\metre} coupled to a superconducting microwave resonator (Fig.~\ref{fig:electrooptic}c) has also been observed~\cite{Fan:18}. The planar geometry of this device allowed precise control over the spatial distribution of the microwave electric field, thus improving the microwave-optical overlap. Although $Q_o$ in these structures is lower at $5\times10^5 - 10^6$, this is offset by the decreased microwave losses, with $Q_{\Omega}\approx1.5\times 10^4$. Single sideband operation is enabled by using TE and TM optical modes as pump and signal modes respectively, an approach permitted by use of the $r_{13}$ electro-optic coefficient. As a result, a total efficiency including insertion losses of $\eta = (2.05\pm0.04) \times 10^{-2}$ was demonstrated, with the cooperativity being $G_0=(0.075 \pm 0.001)$.

Alternative electro-optic schemes have been proposed. Ref.~\cite{Qasymeh:19} suggests the use of graphene as a non-linear medium. In Ref.~\cite{PhysRevA.96.043808}, a device is described in which two strongly coupled optical resonators support a frequency doublet. The coupling is tunable, and so the splitting of the doublet can be selected to match the required microwave frequency. With the optical pump at the lower of the doublet frequencies, the Stokes process is suppressed. Furthermore, because the FSR no longer has to match $\Omega$, there is more freedom over the resonators' size, allowing for smaller optical mode volumes and the potential for larger microwave-to-optical coupling. For realistic device parameters, efficiencies of $\eta=0.25$ should be achievable.

\subsection{Magneto-optically mediated coupling}

\begin{figure}
\includegraphics{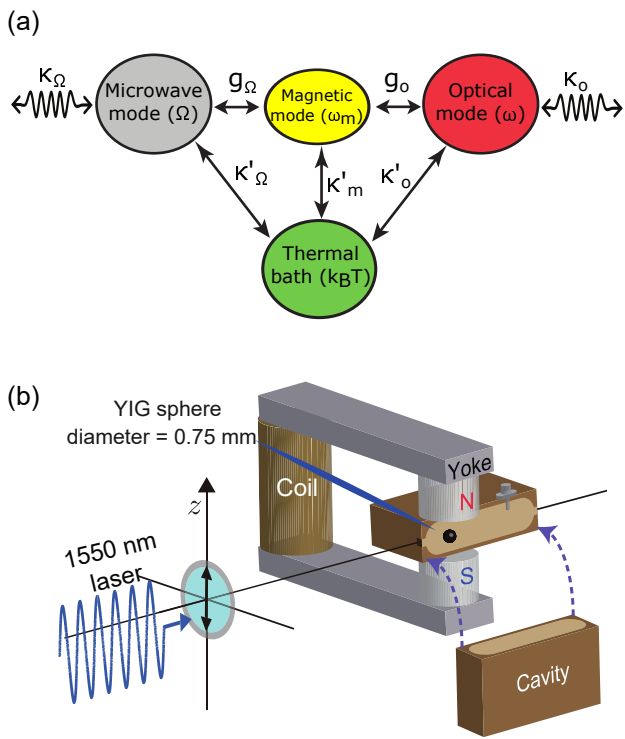}
\caption{Magneto-optic upconversion. (a) Cartoon of process, showing coupling between a microwave and optical mode mediated by a magnetostatic mode, and noise introduced by a thermal bath. (b) An example realization using a microwave frequency copper cavity positioned between the pole pieces of an electromagnet, from Ref.~\protect{\cite{hisatomi}}. In this case, the optical field was not resonantly enhanced.}\label{fig:magnetooptic}
\end{figure} 

In magnetically ordered materials, the Faraday effect results in light traveling in the direction of the magnetization having its plane of polarization rotated by an angle $\theta=\mathcal{V} B$ per unit distance traveled. Here, $B$ is the magnetic flux density in the propagation direction and $\mathcal{V}$ is the material dependent Verdet constant. The effect is non-reciprocal and is used in optical isolators, ring lasers and Faraday rotator mirrors.

In these devices, the magnetization of the material is fixed and $B$ has no time dependency. However, it is also possible to create a time dependent magnetization at microwave frequencies by driving the precession of the magnetization vector about the magnetic field. At small precession angles, the resulting spin waves and magnetostatic modes can be described as weakly interacting bosons termed magnons. As a result of the Faraday effect,  Brillouin scattering between photons of wavevector $k_i$ and magnons of wavevector $q$ can occur, resulting in photons of wavevector $k_o=k_i\pm q$ where the positive sign refers to absorption (an anti-Stokes process) and the negative sign to emission (a Stokes process) of a magnon. The frequencies are related by
\begin{equation}
\omega_o=\omega_i\pm \Omega,
\end{equation}
where $\omega_0$ ($\omega_i$) are again frequencies of two optical modes and $\Omega$ a microwave frequency, and SFG and DFG result.

In order to mediate microwave up-conversion, an electromagnetic signal at frequency $\Omega$ must drive spin waves or magnetostatic modes in a material, thereby creating a coherent magnon population (Fig.~\ref{fig:magnetooptic}). The frequency of the magnons can be tuned to $\Omega$ by varying an applied magnetic field, and the coupling between the two is mediated by magnetic induction. 

The principle material system for quantum magneto-optic experiments has been the ferrimagnetic insulator yttrium iron garnet (YIG). Spin waves in YIG have a narrow linewidth due to low Gilbert damping in the material, leading to its longstanding use as a component in classical filters and transducers~\cite{Gurevich63}. It is also transparent at telecoms wavelengths, with a low absorption constant and a relatively large Verdet constant of \SI{0.008}{\degree\per\gauss\per\centi\metre} at a wavelength of \SI{1.15}{\micro\metre}~\cite{Pisarev1971}. Highly polished YIG spheres with a uniform static magnetization support a well-understood family of magnetostatic modes~\cite{PhysRev.114.739,PhysRev.105.390} with magnetic field dependent frequencies. They also allow WGMs to propagate around the equator~\cite{PhysRevLett.117.123605}, with an optical $Q$ $>10^5$ at \SI{1550}{nm}; this is usually limited by surface scattering. Typically, the diameter of the spheres used are $0.2-\SI{1}{\milli\metre}$, leading to an FSR of $16 - \SI{80}{\giga\hertz}$. The evanescent coupling conditions that apply for WGMs in disks must also be met here, and this can be done by either fiber coupling~\cite{osada}, or prism coupling using rutile~\cite{PhysRevA.92.063845,PhysRevLett.117.133602} or silicon~\cite{PhysRevLett.120.133602} prisms.

There is also a significant impedance mismatch between a 50 $\Omega$ microwave feedline and the magnetostatic modes. This can be reduced by embedding the YIG in a microwave cavity~\cite{PhysRevLett.111.127003,PhysRevLett.113.156401,PhysRevApplied.2.054002,PhysRevLett.113.083603}. The single-spin coupling rate to the microwave mode is given by $g_0=\zeta_e B_0/\sqrt{2}$, where the $\zeta_e$ is the electron gyromagnetic ratio, $B_0=\sqrt{\mu_0\hbar\omega_c/2V_c}$ is the zero-point amplitude of the magnetic field in the mode (with $V_c$ being the effective cavity mode volume, see Section \ref{cavityproperties}, and $\mu_0$ the permeability of free space), and the factor $1/\sqrt{2}$ is due to the fact that only the component of the cavity field co-rotating with the magnetization contributes to the coupling. The magnetostatic mode is a collective spin excitation, and the presence of $N$ spins enhances the coupling~\cite{TABUCHI2016729} to the magnetostatic mode such that $g=\sqrt{N}g_0$. For a magnetostatic mode in a spatially varying cavity mode field,
\begin{equation}
g=\frac{\phi}{2}\zeta_e\sqrt{\frac{\hbar\omega_c \mu_0 \epsilon_r}{V_c}}\sqrt{2 N s},
\end{equation}
where $\omega_c$ is the cavity mode frequency, $N$ is the total number of spin sites comprising the magnetostatic mode, $s$ is the spin per site, and $\epsilon_r$ is the relative permittivity of the dielectric within the cavity. $\phi$ is the overlap between magnetostatic and electromagnetic modes, and is given by
\begin{equation}
\phi=\left| \frac{1}{H_\textrm{max}M_\textrm{max} V_m} \times \int dV (\mathbf{H}\cdot\mathbf{M})\right|.
\end{equation}
Here, $\mathbf{H}$ is the magnetic component of the microwave field and $\mathbf{M}$ is the complex time dependent off z-axis magnetization for the magnetostatic mode. $H_\textrm{max}$ and $M_\textrm{max}$ are the maximum magnitudes of these, and $V_m$ is the spatial volume of the magnetostatic mode.

The engineering requirements to increase photon-magnon coupling are therefore similar to those for enhancing the multiphoton cooperativity: the mode volume should be decreased, the overlap between magnetostatic mode and photon mode increased, and the material chosen to maximize the spin density. Strong coupling between magnetostatic modes and cavities has been demonstrated~\cite{PhysRevLett.111.127003, PhysRevLett.113.156401, PhysRevLett.113.083603}, but the cavity must be resonant with the optical FSR, and it is not always straightforward to realize a microwave cavity with a wide tuning range~\cite{Osborn07,doi:10.1063/1.2929367}.

Conservation of angular momentum requires that the output field resulting from magnon annihilation has the opposite optical polarization to the pump~\cite{PhysRevB.96.094412}. This lifts the requirement for the FSR of a single mode family to match the microwave drive and instead allows the freedom to choose an operating point at which the frequency difference between TE and TM modes is the same as $\Omega$. Furthermore, it leads to intrinsically single-sideband operation, allowing either up-conversion or down-conversion to be selected.

Up-conversion efficiency in current experiments is low, with best results of $\eta \sim 10^{-10}$~\cite{hisatomi}. The inclusion of optical cavity modes in the form of WGMs has not led to a significant improvements~\cite{PhysRevLett.117.123605,osada}, principally because of the small overlap between the spatially uniform Kittel mode, which occupies the entire YIG sphere, and the WGMs, which are confined to the equator, giving a small magnon-photon coupling. Possible routes to improve this include using higher order magnetostatic modes, which are concentrated near the surface of the sphere~\cite{Lambert2015,Morris2017,PhysRevB.97.214423,PhysRevB.96.094412,PhysRevLett.120.133602,Osada_2018} (although these are harder to excite with microwaves), or use a ferromagnetic disc or oblate spheroid (although this is likely to have negative consequences for the linewidth of the magnetic modes.)

%

\subsection{\texorpdfstring{$\Lambda$}{Lambda}-systems and Rydberg atoms}

\begin{figure*}
\includegraphics{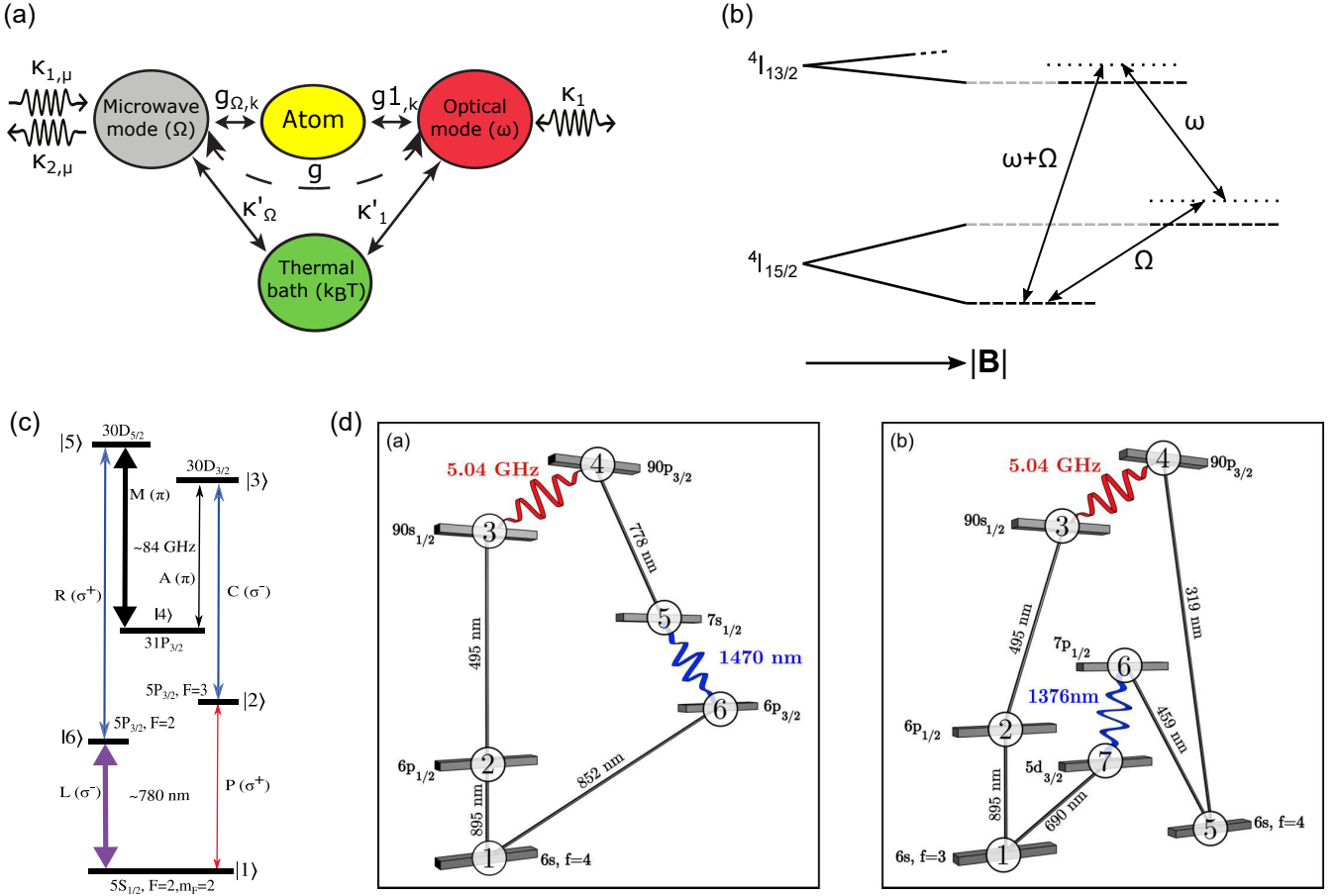}
\caption{Up-conversion using $\Lambda$-systems and similar level structures. (a) Cartoon of process, showing coupling between a microwave and optical mode mediated by atomic levels, and noise introduced by a thermal bath. (b) Schematic of tunable three level system due to an applied magnetic field~\protect{\cite{williamson_magneto-optic_2014}}. (c) A possible set of energy levels in Rb, from Ref.~\protect{\cite{PhysRevLett.120.093201}}. Coherent fields at $\omega$ and $\Omega$ couple the transition at frequency $\Omega + \omega$. (d) Two possible set of energy levels in Cs, from Ref.~\protect{\cite{PhysRevA.96.013833}}. The ability to choose appropriate levels gives flexibility in the output wavelength.}\label{fig:ions_rydberg}
\end{figure*} 

Strong optical non-linearities are also found close to the absorption lines in a medium. For any three levels, the optical transition between one pair of levels must be dark, due to parity selection rules. If the system's dark transition is between the lowest two energy levels, it is termed a $\Lambda$-system~\cite{VogelWelsch}. Coherent Raman scattering in such a set of states can allow for controllable non-linearity, leading to the demonstration of electromagnetically induced transparency~\cite{Harris1997,Marangos98}, slow light~\cite{Hau1999,PhysRevLett.83.1767} and storage of light~\cite{Liu2001,Bajcsy2003}. It also permits microwave up-conversion by three wave mixing, in which a coherent optical pump and microwave signal drive two of the transitions. This results in a coherence on the third transition, which generates an optical field blue shifted by $\Omega$ (Fig.~\ref{fig:ions_rydberg}a).

Erbium dopants in a yttrium orthosilicate crystal (Er$^{3+}$:Y$_2$SiO$_5$) are a good choice for a material system~\cite{williamson_magneto-optic_2014,PhysRevLett.113.063603,fernandez-gonzalvo_frequency_2015}, having a family of optical transitions with narrow linewidths. To create the desired optical transitions, the states $^4I_{15/2}$ and $^4I_{13/2}$ are Zeeman split by the presence of an external magnetic field (Fig.~\ref{fig:ions_rydberg}b). The size of the splitting is proportional to the material's Land\'e $g$-factor. In Er$^{3+}$:Y$_2$SiO$_5$, the large $g$-factor allows for relatively small fields to be used, with a field of \SI{178}{\milli\tesla} giving a splitting of \SI{4.9}{\giga\hertz}. A two photon process involving a microwave photon and an optical photon drives transitions between the parallel spin $^4I_{15/2}$ and the  parallel $^4I_{13/2}$ states, via the antiparallel $^4I_{15/2}$ state. By detuning the microwave and optical fields from resonance by $\sim\SI{10}{\mega\hertz}$, relaxation via spontaneous emission from higher energy states is suppressed, absorption of the signal is avoided and analysis of the dynamics of the problem simplified.

By locating the Er$^{3+}$:Y$_2$SiO$_5$ in a loop-gap microwave resonator ($Q_\Omega\sim300$) the strength of the microwave field can be enhanced. The photon number conversion efficiency for such a scheme is given by~\cite{williamson_magneto-optic_2014,xavithesis} 
\begin{equation}
\label{rare_earth_efficiency}
\eta=\left \lvert \frac{4 i S \sqrt{\kappa_1\kappa_\Omega}}{4\lvert S \rvert^2 + (\kappa_1-2 i\Delta\Omega)(\kappa_\Omega-2 i \Delta\Omega)}\right \rvert^2,
\end{equation}
where mode labels $1$ and $\Omega$ label the output mode and microwave mode respectively, and $\Delta\Omega$ is the detuning of the microwave field from the microwave mode. S is given by
\begin{equation}
S=\sum_k\frac{\omega^{\phantom{*}}_{r,k} g^{\phantom{*}}_{\Omega,k} g^*_{1,k}}{\delta_{1,k}\delta_{\Omega,k}},
\end{equation}
with the sum running over each rare earth ion in the microwave field. For each ion, $\omega_{r,k}$ is the optical field Rabi frequency, $g_{\Omega,k}$ is the coupling to the microwave mode, $g_{1,k}$ is the coupling to the optical output mode, and $\delta$ is the detuning from the modes.

The magneto-optic cooperativity is related to $S$ by $G_o=4|S|/\kappa_1 \kappa_\Omega$, and by making this substitution, and making the total and external losses explicit, we can write the efficiency as
\begin{equation}
\label{rare_earth_efficiency2}
\eta=\frac{4\kappa_{\textrm{e},1}\kappa_{e,\Omega}}{\kappa_1\kappa_\Omega}\frac{G_0}{\left(1+G_0-\frac{4\Delta\Omega}{\kappa_1\kappa_\Omega}\right)^2+\frac{\Delta\Omega^2}{\kappa_1^2\kappa_\Omega^2}(\kappa_1+\kappa_\Omega)^2}.
\end{equation}
A comparison with Eqn (\ref{conversiontraveling}) immediately reveals the parallels this technique and electro-optic approaches. This is not surprising; the coupling to rare earth dopants and Rydberg systems is mediated via their large electric dipoles.

In Ref.~\cite{fernandez-gonzalvo_frequency_2015} an efficiency of $\eta \sim 10^{-12}$ was demonstrated, with the bandwidth of the conversion limited to less than \SI{200}{\kilo\hertz} by the total linewidth of the microwave system. Improvements could be achieved by improving the currently modest quality factor ($Q \approx 300$) of the microwave cavity, perhaps by moving to a superconducting system. More significantly, this demonstration did not use an optical cavity; a doubly resonant cavity (for both pump and signal frequencies) would improve the device efficiency by factor equal to the finesse of the cavity squared, due to the higher LDOS increasing the emission rate.

Proposals for a suitable geometry of an optical cavity include Fabry–P\'erot resonator~\cite{williamson_magneto-optic_2014}, which could be readily integrated into such a scheme, and WGM resonators, as used for electro-optic approaches~\cite{strekalov_microwave_2009}. An alternative way to enhance the optical fields was demonstrated by Zhong et al.~\cite{Zhong2015}, in which a photonic crystal cavity was fabricated directly in neodymium-doped YSO without increasing the inhomogeneous linewidth of the ions in the cavity. Purcell enhancement and dipole-induced transparency were demonstrated.

Ensembles of cold trapped Rydberg atoms~\cite{RevModPhys.82.2313} also exhibit a useful level structure, large dipole moments~\cite{Weller:12}, and a giant electro-optic effect~\cite{Mohapatra08}, allowing microwave-to-optical conversion~\cite{PhysRevA.85.020302}. Specific proposals have been made for caesium~\cite{PhysRevA.96.013833}, rubidium~\cite{Feizpour:16,Kiffner_2016} (Figs~\ref{fig:ions_rydberg}c and \ref{fig:ions_rydberg}d) and ytterbium~\cite{Covey2019} gases to be used, with several pumps coupling multiple transitions allowing an appropriate signal and output wavelength to be selected. Collective states in an gas of Rb have been strongly coupled to superconducting transmission line cavities~\cite{PhysRevLett.103.043603} and up-conversion was first demonstrated~\cite{PhysRevLett.120.093201} using Rydberg states in Rb. Both of these experiments were carried out in the classical regime, with a significant thermal environment, and a maximum  conversion efficiency of $\eta=3\times10^{-3}$ was demonstrated. By ensuring all waves propagate along the same axis, the efficiency was subsequently improved~\cite{PhysRevA.99.023832} to $\eta=0.05$, despite the absence of resonant enhancement of the microwave field. An all-resonant system may be able to achieve $\eta =0.7$. However, because atomic ensembles naturally offer high cooperativity in the phase-matched direction, an optical cavity is not vital for high efficiencies~\cite{Petrosyan19}. This is in contrast to proposals using single atoms~\cite{PhysRevA.96.013833}, which would require resonant enhancement.

\subsection{Optomechanically mediated coupling}

\begin{figure}
\includegraphics{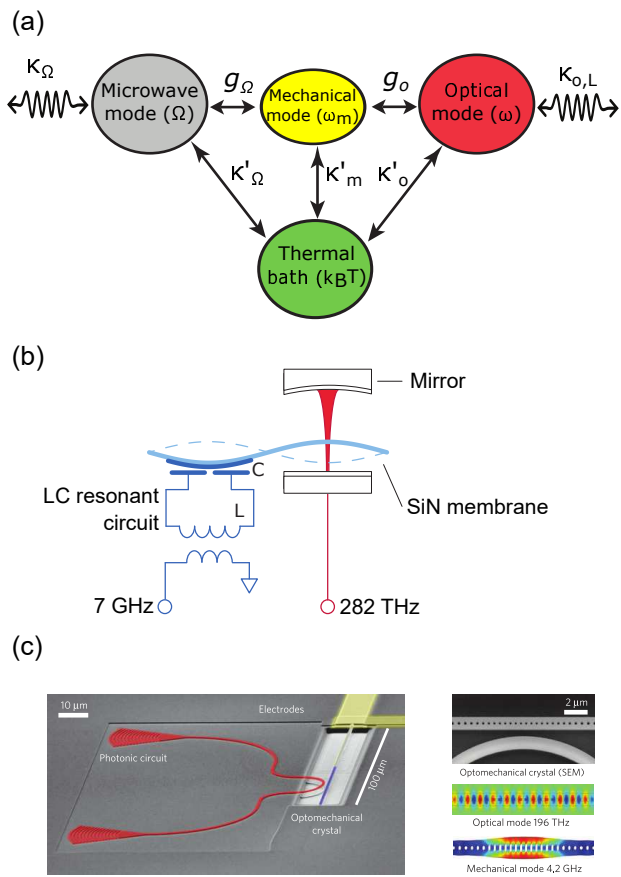}
\caption{Optomechanical upconversion. (a) Cartoon of process, showing coupling between optical and microwave modes mediated by a micromechanical resonator, and noise introduced by a thermal bath. (b) Realization using a vibrational mode in a micromechanical membrane, after Ref.~\protect{\cite{andrews_bidirectional_2014}}. (c) Realization using a photonic crystal cavity in a nanobeam supporting a mechanical breathing mode (right panel), coupled in to a photonic waveguide (left panel), from Ref.~\protect{\cite{bochmann_nanomechanical_2013}}.}\label{fig:optomechanical_scheme}
\end{figure} 

Schemes have been proposed and demonstrated in which microwave and optical fields are coupled via a mechanical resonator~\cite{regal_cavity_2011,Parkins_1999,safavi-naeini_proposal_2011}. Such a device relies upon vibrational modes being  simultaneously coupled to the microwave field via electro-mechanical coupling, and an optical cavity via opto-mechanical coupling~\cite{RevModPhys.86.1391}. If both optomechanical and electromechanical couplings are tunable, then a `swap' operation transfers the excitation from the microwave field to a phonon excitation, and then a second such operation effects a transfer from the mechanical mode to the optical mode~\cite{PhysRevA.68.013808,PhysRevA.82.053806,PhysRevLett.107.133601}. Such a resonant approach requires low levels of dissipation in the photon and phonon modes~\cite{PhysRevLett.108.153603}.

In contrast, coupling through `mechanically dark' modes~\cite{PhysRevLett.108.153603,PhysRevLett.108.153604} has more relaxed requirements. These modes only involve excitations of the optical and microwave cavity modes, and so dissipation due to the mechanical resonator is not present. An analogy can be drawn with dispersive coupling of qubits in a cavity~\cite{Majer07}, for which cavity mode losses are not deleterious.

Coupling through dark modes has been realised by Andrews et al.~\cite{andrews_bidirectional_2014} and Bagci et al.~\cite{bagci_optical_2014}. In both of these studies, vibrational modes with frequencies between \SI{340}{\kilo\hertz} and \SI{1240}{\kilo\hertz} with $Q\sim10^6$ supported by a silicon nitride membrane with characteristic lengths $\sim\SI{100}{\micro\metre}$ were used. The membrane was positioned on the axis of an optical cavity such that a change in position of the membrane changes the length of the cavity and hence the mode frequencies. The microwave resonator was an inductor-capacitor (LC) circuit  with a capacitative component close to the membrane. Coupling was mediated by coating part of the membrane with superconducting niobium~\cite{andrews_bidirectional_2014}, or with aluminium~\cite{bagci_optical_2014}, such that a change in membrane position modulated the capacitance of the circuit and therefore changed its frequency. Applying strong pumps to each cavity, red-detuned from resonance, enhanced the photon-phonon couplings. Conversion efficiencies of $\eta = 8\times10^{-3}$~\cite{bagci_optical_2014} and $\eta = (8.6\pm0.007)\times10^{-2}$~\cite{andrews_bidirectional_2014} were demonstrated.

The narrow linewidth and low frequency of the intermediate mechanical mode results in both a small bandwidth and residual thermal occupancy at 10 mK. However, by exploiting the correlations between the output noises from the microwave and optical cavities, and using a classical feed-forward protocol, Higginbotham et al.~\cite{Higginbotham2018} reduced the thermal noise added during up-conversion, at the expense of added noise from the feed-forward process. Furthermore, improvements to the coupling and mode matching allowed an efficiency of $\eta = 0.47\pm0.001$ to be achieved.

Alternatively, the optical mode can be confined  in the nanomechanical element itself~\cite{Bochmann2013,Vainsencher:16}, with a piezoelectric coupling between the two excitations. In these experiments, the optical field is evanescently coupled from a rib waveguide into a photonic crystal cavity fabricated in a nanobeam resonator. The nanobeams have a breathing mode ($\sim\SI{4}{\giga\hertz}$), with $Q \sim 1000$. This relatively high frequency allows suppression of thermal phonons at dilution fridge temperatures, and allows resonant driving by a microwave field. An internal efficiency of $\eta = (4\pm2)\times 10^{-3}$ and an external efficiency of $\eta = (1.4\pm0.6)\times 10^{-4}$ was demonstrated~\cite{Vainsencher:16}.

Finally, we note that optomechanical approaches are intrinsically bidirectional, because of the symmetry between the optical and microwave cavity fields in the Hamiltonian. In several cases~\cite{andrews_bidirectional_2014, Vainsencher:16}, optical-to-microwave conversion was demonstrated.

\section{Conclusion and future perspectives}

%
%

\begin{figure*}
\centering

\begin{tikzpicture}
\begin{semilogyaxis}[
    grid=both,
		grid style={line width=0.1pt, draw=gray!10},major grid style={line width=0.2pt,draw=gray!50},
		width=16cm,height=8cm,
    xlabel={Year},
    xmin=2006,xmax=2020,
		ytick={1,1e-2,1e-4,1e-6,1e-8,1e-10,1e-12},
		yticklabels={$1$,$10^{-2}$,$10^{-4}$,$10^{-6}$,$10^{-8}$,$10^{-10}$,$10^{-12}$},
    ylabel={Efficiency, $\eta$},
    ymin=1e-14,ymax=1.4,
    legend pos=north west,
		legend cell align={left},
		/pgf/number format/.cd,
        use comma,
        1000 sep={}
]

\addplot[
		only marks,
    color=blue,
    mark=square,
		nodes near coords={\cite{\thecitation}},
    nodes near coords align={\alignment},
    visualization depends on={value \thisrow{cite} \as \thecitation},
		visualization depends on={value \thisrow{align} \as \alignment},
    every node near coord/.style={black}
    ]
		table[x=x, y=y]{eo.dat};
		
    \addlegendentry{Electro-optic}

\addplot[
		only marks,
    color=green,
    mark=triangle,
		nodes near coords={\cite{\thecitation}},
    nodes near coords align={right},
    visualization depends on={value \thisrow{cite} \as \thecitation},
    every node near coord/.style={black}
    ]
		table[x=x, y=y]{mo.dat};
		
    \addlegendentry{Magneto-optic}	
		
\addplot[
		only marks,
    color=orange,
    mark=diamond,
		nodes near coords={\cite{\thecitation}},
    nodes near coords align={\alignment},
    visualization depends on={value \thisrow{cite} \as \thecitation},
		visualization depends on={value \thisrow{align} \as \alignment},
    every node near coord/.style={black}
    ]
		table[x=x, y=y]{lambda.dat};
		
    \addlegendentry{$\Lambda$- and Rydberg systems}
		
\addplot[
		only marks,
    color=red,
    mark=*,
		nodes near coords={\cite{\thecitation}},
    nodes near coords align={right},
    visualization depends on={value \thisrow{cite} \as \thecitation},
    every node near coord/.style={black}
    ]
		table[x=x, y=y]{om.dat};
		
    \addlegendentry{Optomechanical}

\end{semilogyaxis}
\end{tikzpicture}

\caption{Progress in microwave up-conversion efficiency. Electro-optic techniques (blue squares) were initially applied to non-resonant systems~\protect{\cite{Khan:07}} before resonant enhancement of first the optical field~\protect{\cite{savchenkov_tunable_2009, strekalov_efficient_2009}}, and then also the microwave field~\protect{\cite{Rueda:16, Fan:18}}. Recent maneto-optical upconversion experiments (green triangles) using WGMs in YIG~\protect{\cite{hisatomi, osada, PhysRevLett.117.123605}}, while showing promise, still suffer from low efficiencies, although in future this may increase with the use of resonant microwave fields. Current demonstrations based on rare earth ions~\protect{\cite{fernandez-gonzalvo_frequency_2015}} (orange diamonds) are also low in efficiency, but this figure was achieved without an optical cavity. More recently, higher efficiencies have been achieved using Rydberg states in rubidium atoms~\protect{\cite{PhysRevLett.120.093201, PhysRevA.99.023832}}. Optomechanical systems (red circles) have shown the highest demonstrated efficiencies so far~\protect{\cite{andrews_bidirectional_2014, bagci_optical_2014, Vainsencher:16}}, but require laser cooling or feed-forward techniques to suppress thermal noise~\protect{\cite{Higginbotham2018}}. We also show the efficiency achievable by using a commercial electro-optic modulator~\protect{\cite{ThorLabs}}.} \label{fig:efficiency_progress}
\end{figure*}
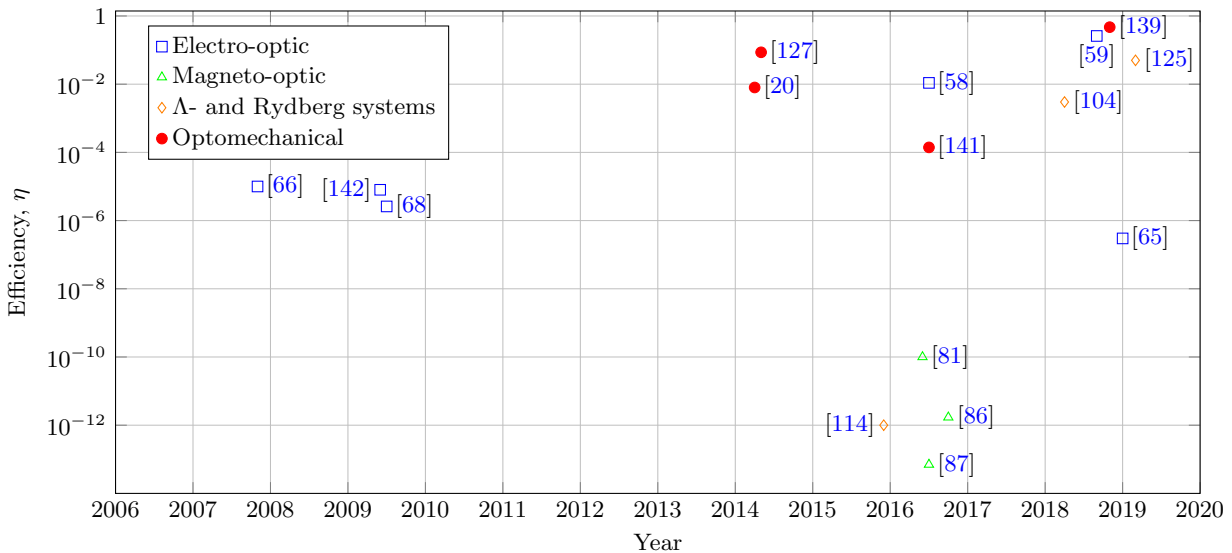

{
\setlength{\tabcolsep}{1em}

\begin{table*}
\caption{Figures of merit for selected experiments for microwave to optical frequency up-conversion.  If the loss rates are not given explicitly, we assume critical coupling and 1 mW of optical pump power.\\}
\begin{tabular}{ c c c c c c }
\hline
Ref.&Type&Material system&$\Omega$ (GHz)&$\eta$&Bandwidth (MHz) \\
\hline
\hline
\cite{Rueda:16}&Electro-optic&LiNbO$_3$&8.9&$0.0109$&1.38\\ 

\cite{xavi2}&$\Lambda$-system&ER:YSO&5.18&$1.2\times10^{-5}$&0.13\\

\cite{Fan:18}&Electro-optic&AlN&8.31&0.259&0.59\\

\cite{hisatomi}&Magneto-optic&YIG&10.5&$10^{-10}$&2.28\\

\cite{PhysRevA.99.023832}&$\Lambda$-system&Rubidium atoms&84&$5\times10^{-2}$&15\\

\cite{Higginbotham2018}&Optomechanical&Si$_3$N$_4$&6.16&$0.47$&$1.2\times10^{-2}$\\ 

\hline
\end{tabular}
	
	\label{efficiencies}
\end{table*}

}

The importance and intricacies of the challenge of coherent microwave up-conversion has led to a diverse range of experimental techniques being brought to bear upon it, each with their own advantages and drawbacks. Maximizing the internal efficiency of the device has been an important goal, and the current record, achieved in an optomechanical system~\cite{Higginbotham2018}, is set at $\eta=0.47$. For approaches involving nonlinear media (including hosts for $\Lambda$-systems) embedded in cavities, the principle roots to maximising efficiency are increasing the non-linearity of the medium, increasing the $Q$ factor of the cavities, and increasing the overlap and confinement of the three electromagnetic modes. These criteria are not independent, and often a compromise between them is necessary.

It is in this area of optimal cavity mode forms that current magneto-optic schemes suffer; the overlap between the Kittel mode, which occupies the whole sphere, and the whispering gallery modes is small. Higher order magnetostatic modes, which are concentrated on the surface of the ferromagnet, may offer a route for improvement, although efficient coupling into such modes is difficult, requiring a highly non-uniform driving field. Alternatives may include using magnetic modes in rare earth crystals~\cite{arXiv:1812.03246}, hybrid magneto-optomechanical devices~\cite{arXiv:1904.07779}, and travelling spin waves in thin films~\cite{Kostylev2019}.

For the relatively low efficiency up-conversion demonstration~\cite{fernandez-gonzalvo_frequency_2015} using Er:YSO, the optical field was not confined. Calculations suggest an improvement of $\sim10^{10}$ are possible with an optimized cavity. Electro-optic experiments have demonstrated the benefits of double cavities~\cite{Rueda:16}, with further room for improvement of the relatively low $Q$ microwave cavities. Very often, however, increased efficiency due to higher $Q$ cavities is bought at the expense of conversion bandwidth. In the case of efficient optomechanical up-conversion demonstrations, bandwidth has typically been limited by the mechanical resonator. There is some leeway to tune the bandwidth by overcoupling of cavities.

Noise in the form of additional microwave photons can be added to the output by either inadvertent down conversion of optical photons to microwave frequencies (DFG), or by thermal processes in cavities with low resonant frequencies, and therefore significant thermal occupancy. All the approaches described here admit techniques to suppress or avoid DFG in various ways, principally by decreasing the LDOS at $\omega-\Omega$. In electro-optic systems, this is done by detuning of the optical pump from resonance~\cite{strekalov_efficient_2009}, engineering of the optical FSR by using anticrossings with modes of different polarisation~\cite{Rueda:16}, or exploiting off-diagonal elements of the electro-optic coefficient to up-convert to the opposite polarisation~\cite{Fan:18}. Additionally, selection rules for magnon scattering further suppress DFG for magneto-optic systems. $\Lambda$-systems, on the other hand, do not exhibit DFG processes; nor do systems in which coupled resonators form a frequency doublet.

At temperatures less than \SI{100}{\milli\kelvin}, the thermal occupancy of microwave frequency cavity modes is much less than one photon. But some optomechanical systems rely on intermediate mechanical modes with lower frequencies and they still have a significant thermal population at these temperatures, resulting in up-conversion of thermal photons. This can be reduced by either laser cooling of the mechanical membrane or an active feed forward protocol~\cite{Higginbotham2018}. A recent experiment~\cite{arXiv:1812.07588} has suppressed added noise in an optomechanical system by raising the operating frequency of the mechanical mode.

In conclusion, advances in microwave up-conversion have been rapid. The relevant engineering figures of merit are quantum efficiency, bandwidth and fidelity. Progress on efficiency has been significant, with best efficiencies improving from $\eta=10^{-5}$ to $\eta=0.47$. Often, however, efficiency is traded against bandwidth, with high $Q$ cavities increasing interaction strengths at the cost of a narrower linewidth. Direct measurements of fidelity are less common, but noise measurements have been performed on optomechanical systems.

The breadth of techniques brought to bear on the problem have made it a fruitful area of study; significant progress has been made on such diverse topics as Rydberg atoms, dilute spin ensembles, control of nanomechanical oscillators and non-linear magneto- and electro-optics. The path to reach efficiencies close to unity will also prove to be a rich seam of physics.


\begin{acknowledgments}

We thank Amita Deb for useful comments on this manuscript. We acknowledge support from the MBIE Endeavour Smart Ideas fund. 

\end{acknowledgments}

\bibliographystyle{ieeetr}

\bibliography{firstbib}

\end{document}